\documentclass[]{aa} 
\usepackage{graphicx}
\usepackage{amsmath}
\usepackage{txfonts}
\usepackage{natbib}
\usepackage{xcolor} 
\usepackage{adjustbox} 
\usepackage{tabularx} 
\usepackage{hyperref}
\hypersetup{
        colorlinks=true,
        linkcolor=blue,
        filecolor=blue,      
        urlcolor=blue,
        citecolor=blue
}

\bibpunct{(}{)}{;}{a}{}{,}

\begin{document}
   
   \title{ALMA small-scale features in the quiet Sun and active regions}

   \author{R. Braj\v{s}a   \inst{1}
          \and
          I. Skoki\'{c}   \inst{1}
          \and
          D. Sudar  \inst{1}
          \and
          A. O. Benz \inst{2, 3}
          \and 
          S. Krucker \inst{2, 4}
          \and 
          H.-G. Ludwig \inst{5}
          \and 
          S. H. Saar \inst{6}
          \and 
          C. L. Selhorst \inst{7}
         }

   \institute{Hvar Observatory, Faculty of Geodesy, University of Zagreb,
              Ka\v{c}i\'{c}eva 26,  10000 Zagreb, Croatia
                 \and
              University of Applied Sciences and Arts Northwestern Switzerland, Bahnhofstrasse 6, 5210 Windisch, Switzerland
              \and
              Institute for Particle Physics and Astrophysics, ETH Zurich, 8093 Zurich, Switzerland
              \and
              Space Sciences Laboratory, University of California, Berkeley, CA 94720-7450, USA
              \and 
              Landessternwarte, Zentrum f\"ur Astronomie der Universit\"at Heidelberg, K\"onigstuhl 12, 69117 Heidelberg, Germany 
              \and 
              Harvard-Smithsonian Center for Astrophysics, 60 Garden Street, Cambridge, MA 02138, USA             
              \and
              NAT - N\'ucleo de Astrof\'isica, Universidade Cruzeiro do Sul/Universidade Cidade de S\~ao Paulo, S{\~a}o Paulo, SP, Brazil              
                                         }

\offprints{Roman Braj\v sa, \email roman.brajsa@geof.unizg.hr}

   \date{Release \today}

\abstract
{}
{The main aim of the present analysis is to decipher (i) the small-scale bright features 
in solar  images of the quiet Sun and active regions obtained with the Atacama Large Millimeter/submillimeter Array (ALMA) 
and (ii) the ALMA correspondence of various known chromospheric structures visible in the 
H$\alpha$ images of the Sun.}
{
Small-scale ALMA bright features in the quiet Sun region were
analyzed using single-dish ALMA observations (1.21 mm, 248 GHz) and 
in an active region using interferometric ALMA measurements (3 mm, 100 GHz). 
With the  single-dish observations, a full-disk solar image is produced, while interferometric measurements enable the
high-resolution reconstruction of  part of the solar disk, including the active region. 
The selected quiet Sun and active regions are compared with the H$\alpha$ (core and wing sum), EUV, and soft X-ray images and with 
the magnetograms. 
}
{ 
In the quiet Sun region, enhanced emission seen in the ALMA is almost always associated with a strong line-of-sight (LOS)
magnetic field. 
Four coronal bright points were identified, while other small-scale ALMA bright features are most likely associated with magnetic network elements and plages. 
In the active region, in 14 small-scale ALMA bright features randomly selected and compared with other images, we found 
five good candidates for coronal bright points, 
two for plages, and five for fibrils. Two unclear cases remain: a fibril or a jet, and a coronal bright point or a plage.
A comparison of the H$\alpha$ core image and the 3 mm ALMA image of the analyzed active region 
showed that the sunspot appears dark in both images (with a local ALMA radiation enhancement in sunspot umbra), the four 
plage areas are bright in both images and dark small H$\alpha$ filaments are clearly recognized as dark structures of the same shape also 
in ALMA. 
}
{}


\authorrunning{R. Braj\v sa et al.}
\titlerunning{ALMA small-scale features in the quiet Sun and active regions}

   \maketitle

\section{Introduction}

The solar chromosphere, the main source region for the solar radiation at millimeter and submillimeter wavelengths, 
is not understood to a satisfactory level.
It is dynamically coupled 
with the photosphere below and with the transition region and corona above
\citep{Gabriel1992, Golub1997}.
The Atacama Large Millimeter/submillimeter Array 
(ALMA){\footnote{http://www.almaobservatory.org}$^{\rm ,}$\footnote{https://www.eso.org/sci/facilities/alma.html}}
enables important new observations of this layer of 
the solar atmosphere \citep{Bastian2018, Loukitcheva2019}, 
using both its single-dish \citep{White2017}
and interferometric \citep{Shimojo2017} observing modes. 

The main source of ALMA emission at millimeter wavelengths is thermal bremsstrahlung, that is, electron-ion free-free transitions  \citep{Wedemeyer2016, Brajsa2018ceab}. For $\lambda$ = 1 mm, the contribution function includes the heights in the 400 km to 1400 km range, 
with the maximum at $h$ = 700 km,  while for $\lambda$ = 3 mm the contribution function covers the 600 km to 1600 km range, with the 
maximum at $h$ = 950 km \citep{Wedemeyer2016}.

The present analysis relates small-scale ALMA features to known solar structures.
 The main  candidates in our focus are coronal bright points (CBPs), chromospheric plages, magnetic network,  
 H{$\alpha$} fibrils, and filaments. 

We note that CBPs belong to the solar corona, while ALMA maps the solar chromosphere. So, the present analysis refers in fact to the 
chromospheric correspondence of CBPs.

The  CBPs are small-scale coronal loops observed with enhanced emission in UV, EUV,
and X-rays (see e.g., the review by \citealt{Madjarska2019}). These loops are always associated with magnetic elements of opposite polarities as seen in magnetogram data. 
About 50\% of CBPs are related to newly emerging flux regions, while others are related to random encounters of magnetic flux \citep{Mou2016, Mou2018}. At chromospheric and transition-region temperatures, it is mainly the footpoints of the CBP loops that are visible \citep{Madjarska2019}. 

In spite of different morphological forms being revealed in EUV images of the Sun (point-like structures, small loops, and small active regions \citep{Brajsa2001, Brajsa2002}), CBPs can be regarded as 
downscaled active regions \citep{Madjarska2019} and may contribute to the coronal heating (e.g., \citealt{Rekowski2006}). 

\citet{Harvey1985} reported that the CBPs are associated with He I 10830\,{\AA} dark points (localized regions of enhanced He I absorption).
That study was followed by comprehensive analyses of CBPs identified at the wavelength of 20 cm with the VLA \citep{Habbal1986, Habbal1988}. 
The main results of previous radio observations of CBPs at cm wavelengths were summarized by \citet{Madjarska2019}. 
Solar observations with ALMA provide an 
extension of these efforts to the smaller wavelengths in the millimeter and submillimeter wavelength ranges.

The motions of CBPs were used as tracers for studying solar rotation including transport 
of angular momentum toward equator \citep{Brajsa2002, Wohl2010, Sudar2015, Sudar2016}
and turbulent diffusion  
\citep{Brajsa2008, Brajsa2015, Skokic2016, Skokic2019}. 
\citet{Brajsa2004} found the height of CBPs in the 8000 - 12000 km range using the Solar and Heliospheric Observatory (SOHO) 
Extreme ultraviolet Imaging Telescope (EIT) data (284\,{\AA} channel), and \citet{Sudar2016} found  
heights of about 6500 km using the Solar Dynamics Observatory (SDO) Atmospheric Imaging Assembly (AIA) data
(193\,{\AA} channel).  These results are in general agreement with the ones obtained by 
\citet{Kwon2010, Kwon2012} using the stereoscopic method.

Localized bright patches in polar regions of the Sun were investigated at 17 GHz using the Nobeyama Radioheliograph (NoRH) by \citet{Selhorst2017}. 
The authors concluded that the increased emission of the 17 GHz bright patches comes from changes in the chromospheric 
plasma, instead of the coronal ones observed in the EUV lines. 
In an earlier study of small bright patches in 17 GHz NoRH maps 
\citep{Nitta2014}, it was not possible to find a correlation between the 17 GHz polar sources and small-scale bright regions in the 
SDO-AIA
images, and it was concluded that the 17 GHz sources might be artifacts created by the image synthesis and 
deconvolution used in the NoRH maps.  This is an additional motivation to devote particular attention to the image reconstruction of 
ALMA data, especially regarding image orientation, co-alignment, and so on.

We now briefly summarize earlier results on CBPs obtained with ALMA. 
\citet{Shimojo_Hudson2017} observed a plasmoid ejection from a CBP in solar active region simultaneously 
at millimeter wavelengths with ALMA, at EUV wavelengths with SDO-AIA, and in soft X-rays with the Hinode X-ray Telescope (XRT). 
The authors 
concluded that 
the plasmoid consists either of approximately isothermal plasma with   $T\approx 10^5$ K 
that is optically thin at 100 GHz, or a core with $T\approx 10^4$ K and a hot envelope. 
The analysis showed the value of the additional temperature and density constraints provided by ALMA. 
\citet{Rodger2019} analyzed the spectrum of 
the plasma eruption using subband data within the ALMA Band 3. 
They  concluded that both stationary and moving enhancements of radiation 
are partially  
optically thick at 100 GHz. 
The electron temperatures were estimated to lie between 7370 K and 15300 K for the stationary elements  
and between 7440 and 9560 K for the moving elements.
Plasmoid ejection from a CBP is further analyzed in the present work.

\citet{Brajsa2018aa} compared the full-disk solar ALMA image at  $\lambda = 1.21$ mm with simultaneous 
H$\alpha$, He I, and EUV images as well as with a magnetogram. 
They found that active regions appear as bright areas in ALMA observations, 
while prominences on
the disk and coronal holes are barely discernible from the quiet Sun background and have a slightly lower intensity than surrounding
quiet Sun regions. Magnetic inversion lines appear as large, elongated dark structures. 
The  CBPs show a high correspondence with ALMA bright features: 
the vast majority (75\%) of all CBPs from the EUV image correspond to
He~I 10830\,{\AA} dark points, and the vast majority (82\%) of all CBPs from the EUV image correspond to the
ALMA 1.21 mm bright points.

Most parts of active region areas are occupied by bright plages \citep{Martres1977, Zirin1988, Zirin1992, Gurman1992}. The
chromospheric plage is an extended emission region observed in strong chromospheric
lines such as H$\alpha$ and He II 304~\AA. These bright regions in the H$\alpha$ core correspond
to longitudinal (vertical) magnetic field peaks of approximately 1000 Gauss and up to
1500 Gauss. Above the plage is a coronal condensation emitting enhanced EUV, X-ray,
and radio emission. These structures are hotter and have a higher density than the
surrounding regions and may form a magnetic canopy \citep{Solanki1993, Solanki2004}. Outside active
regions, plages are usually simply called network elements and consist of unipolar
magnetic regions, where the magnetic network is relatively strong  and of one
polarity. As in active regions, the network is bright in the H$\alpha$ core.

Plage and network magnetic fields are distinguished from each other by the fact
that the magnetic field typically present in a plage is higher. 
Plages are found inside active regions, while network fields are present all over
the Sun (including the decay products of active regions, which form the so-called
enhanced network) but concentrated mainly at the edges of supergranules with a length
scale of 15--30 Mm \citep{Borrero2017}. At photospheric height, where most of the
measurements of solar magnetic fields were taken, both plage regions and the
network elements are composed of groups of more or less discrete magnetic flux
concentrations \citep{Borrero2017}.

Chromospheric fibrils are dark, elongated structures visible in H$\alpha$ (centerline or off band) with a width of 725 -- 2200  km 
and an average length of 11000 km  \citep{Martres1977, Zirin1988}. They mark horizontal lines of force  connecting 
opposite magnetic polarity, and so they may be used to estimate the configuration of the magnetic field in and around active regions. 
Fibrils play an important role in the process of filament formation when fibrils no longer connect the areas of opposite 
polarities but curve into the polarity inversion zone that will form the filament channel   \citep{Tandberg1995}.  
In this way, fibrils play an important role in filament formation and are closely related to them \citep{Aschwanden2019}.

Recently, \citet{Rutten2017II} and \citet{Rutten2017III} performed an extensive new analysis of chromospheric fibrils. 
They introduced three subclasses of fibrils, and the most interesting one is the so-called contrail fibril representing the track of 
the cooling gas along the path where a transient dynamic heating event took place several minutes before. 
Moreover, \citet{Rutten2017III} 
predicted that at the ALMA wavelengths the general appearance of the solar atmosphere would be similar to the H$\alpha$ images
with a good dark-dark correspondence,  but with greater fibril opaquenesses at millimeter wavelengths, which increases with wavelength
and with less lateral fibril contrast due to a lack of sensitivity to Doppler shifts. For the present work, we investigated this hypothesis.

In this paper, we present an analysis of various features observed by ALMA in quiet and active regions on the Sun. 
We raise two main scientific questions. Firstly, what are the small-scale bright features 
in solar ALMA images and particularly in the quiet Sun observed in single-dish mode and in the active region  
using the interferometric mosaic? Secondly, how do the chromospheric structures visible in the 
H$\alpha$ images relate to features of ALMA images?
This paper extends the results of \citet{Brajsa2018aa} by studying,  in detail, several ALMA features in a quiet region and by 
including interferometric observations of an active region.  
The ALMA images were compared with observations taken 
in other wavelength ranges (magnetogram, soft X-rays, UV, EUV, and H$\alpha$) at the same time. After presenting the results of the analysis, we 
discuss the relationship of the identified ALMA features with features observed at other wavelengths regarding their intensities, positions, 
and shapes.


\section{Data and reduction methods}

\begin{figure*}
        \centering
        \includegraphics[width=17cm, trim={0 0.5cm 0 0},clip]{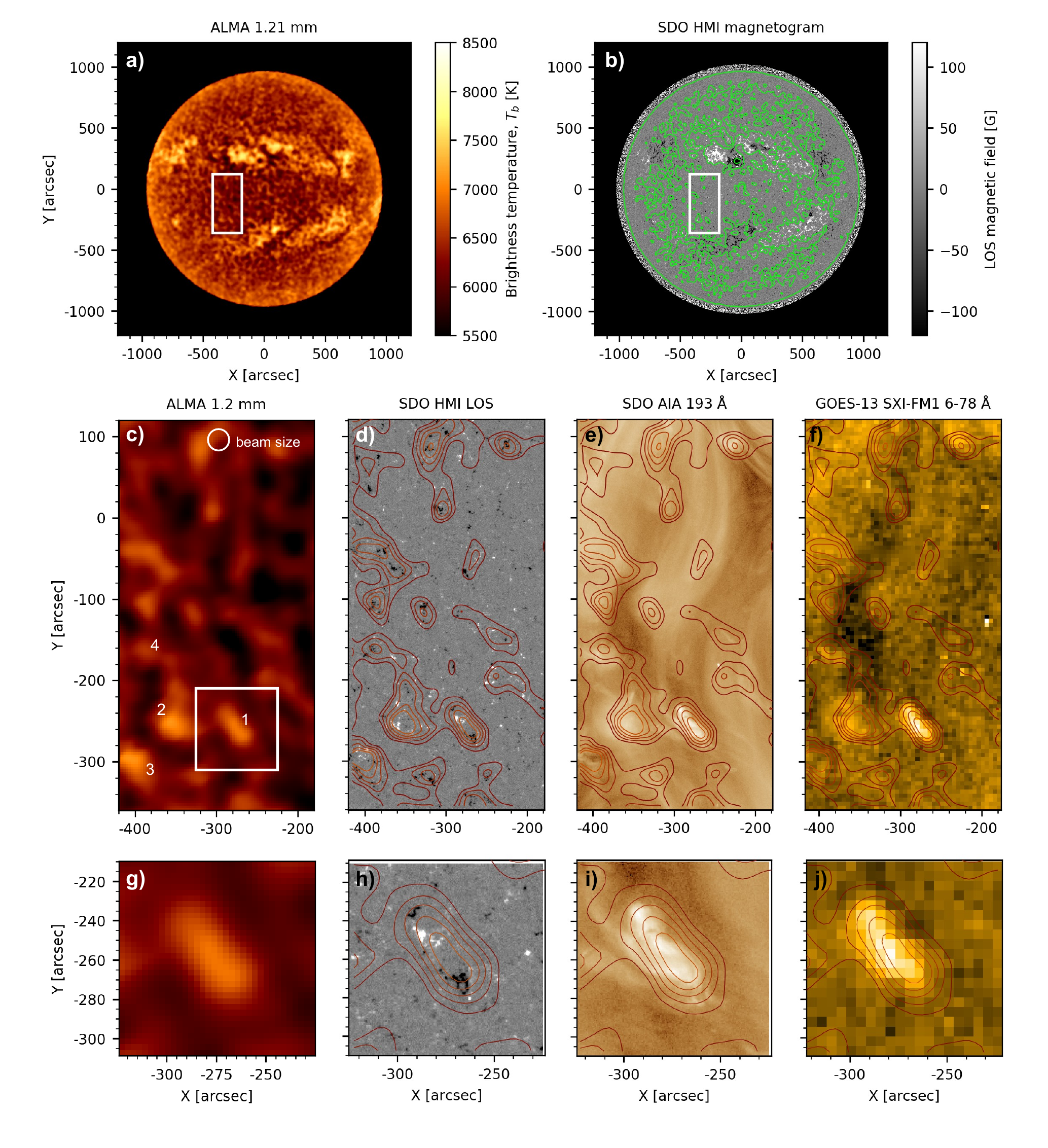}
        \caption{
        (a): Single-dish ALMA 248 GHz (1.21 mm, Band 6).    
        (b): SDO-HMI magnetogram,  
        with white box outlining the quiet Sun area shown in middle row panels. 
        Both images were taken on 18 December 2015 at 20:12 UT. Magnetogram intensity was clipped at $\pm$120 Gauss 
         and ALMA contours of 6500 K were overlaid in green.
        (c):  ALMA image of the quiet Sun shown enlarged and compared with 
        (d): SDO-HMI magnetogram, 
        (e): SDO-AIA 193~\AA~filtergram, and
        (f): GOES-13 SXI tin (Sn) filter image, with ALMA contours overlaid  (levels at 6300, 6400, 6500, 6600, and 6700 K).
        (g-j): Enlargements of a single ALMA bright feature, marked  in the white rectangle in panel (c), 
                at the same wavelengths as in the middle row. 
                The beam size of the  single-dish ALMA measurements, 26", is given in panel (c). 
                }
        \label{Fig_1}%
\end{figure*}

\subsection{ALMA single-dish data}

An image of the whole solar disk from the December 2015 Science Verification
campaign is used in the current analysis. 
The 18 December 2015 
full-disk solar map obtained by scanning the solar disk with a 12 m single dish 
total power ALMA antenna (PM03) at a frequency of 248 GHz corresponding to 
$\lambda = 1.21$ mm in a double circle pattern \citep{Phillips2015, White2017} is presented in Fig.~\ref{Fig_1}a. 
This figure is adapted from \citet{Brajsa2018aa}. 
 The observing wavelength is part of  Band 6. 
We restricted our analysis to that band because on 18 December 2015 
no full-disk image in the other observing band (Band 3) was available.
This observing day was chosen because of the availability of images in other 
wavelengths. 
The observation started at 20:12:21 UT on 18 December 2015 and lasted for 
about 17 minutes including calibration scans \citep{White2017}. 
The resulting image has a single-dish beam size of 26" and is oversampled at 3" per pixel. 

The standard ALMA calibration method \citep{White2017} for solar single-dish images was used, which includes
a correction (multiplication) factor $C$ applied to account for the antenna efficiency ($C = 1.16$ for Band 6).
Data reduction was performed using the Common Astronomy Software Applications (CASA) 
package\footnote{http://casa.nrao.edu}.
The final uncertainty in the brightness temperature is estimated to be about 5-10  \% 
\citep{White2017, Shimojo2017}. The data reduction and analysis of the single-dish data 
are described in more detail by \citet{Brajsa2018aa}.

\subsection{ALMA interferometric data}

We used the Band 3 ($\nu$ = 100 GHz, $\lambda$ = 3.0 mm) image of the active region 12470 from the ALMA Science Verification data. 
The data were taken on 16 December 2015 with a heterogeneous array of 
 twenty-one 12 m antennas and nine 7 m antennas. 
Calibration and imaging were performed in CASA using the procedure described by \citet{Shimojo2017}. The observation was a 149 pointing mosaic with a separation between pointings of 24.1" (corresponding to the Nyquist sampling) and an integration time for each point of 6.048 seconds. All four spectral windows 
(subbands) were combined into a single image for better quality with the final synthesized beam size (spatial resolution) of 
 4.9" $\times$ 2.2". 
The interferometric image was combined with the full-disk single-dish image of the Sun (taken at the same time) using a CASA feather task to recover the absolute flux scale missing from the interferometer data (for details, see \citealt{Shimojo2017}). The field of view is 
 300" $\times$ 300". 
The interferometric image is presented in Fig.~\ref{Fig_2}a.

Details of the interferometric data used here are given in Table 4 in \citet{Shimojo2017}. 
The observation lasted for 47 minutes, and the reference time was 18:32:41 UT, corresponding roughly to the middle of the observation. The reference time is arbitrarily chosen during data calibration and imaging and is used to fix the interferometer visibility data of the moving target (Sun) to the same time, that is, the same position in the sky.

\subsection{SOLIS-FDP H$\alpha$ data}
The Synoptic Optical Long-term Investigations of the Sun (SOLIS) is a synoptic facility  designed to study magnetic and non-magnetic solar activity with high spectral and moderate spatial resolution over a long time frame \citep{Keller2003}.
The Full Disk Patrol (FDP) is a SOLIS full-disk imager with a resolution of 1" per pixel\footnote{https://www.nso.edu/telescopes/nisp/solis/}. We used H$\alpha$ core and wing sum images centered at 6563~\AA, taken between 18:28:18 UT and 18:54:25 UT on 16 December 2015, roughly corresponding to the observation time of the ALMA interferometric image. 
To improve the quality of the image, 70 individual frames were stacked together and averaged. 
Applying this procedure meant that the final image was slightly sharpened.

\subsection{SDO observations}
We used the SDO-AIA  
data\footnote{http://sdo.gsfc.nasa.gov} \citep{Lemen2012}
 at 171\,{\AA}, 193\,{\AA}, 304\,{\AA}, and 1700~\AA, taken at 20:12 UT on 18 December 2015 and 18:32 UT on 16 December 2015. The spatial resolution is 0.6" per pixel in all images. We also used the SDO Helioseismic and Magnetic Imager (HMI) data \citep{Scherrer2012}, which were taken at the same times with the spatial resolution of 0.5" per pixel. 
 In order to better characterize different features observed with ALMA, their temporal evolution and dynamics in the HMI and AIA channels were analyzed using JHelioviewer visualization software \citep{JHelioviewer2017}.
The primary ions, regions of the atmosphere and characteristic temperatures observed by AIA channels used herein are listed in Table \ref{tables:aia}. These selected channels map layers of the solar atmosphere of increasing height and have high count rates in the quiet Sun and active regions \citep{ODwyer2010}.
        
\subsection{GOES observations}  
The Geosynchronous Operational Environmental Satellite 13 (GOES-13) Solar X-ray Imager (SXI) data\footnote{https://sxi.ngdc.noaa.gov} were taken through the tin (Sn) filter with a wavelength range of 6-78~\AA\, and a spatial resolution of 5" per pixel. The image was recorded at 20:12 UT on 18 December 2015.

\begin{table*}
        \caption{SDO/AIA channels, adapted from \citet{Lemen2012}.}
        \label{tab}
        \begin{center}
                \begin{tabular}{cccc}
                        \hline
                        Channel [\AA]& Primary ion(s) & Region of atmosphere &Char. temp. [K]\\
                        \hline
                        1700 & continuum &temperature minimum, photosphere &$5000$\\
                        304 & He II &chromosphere, transition region &$50000$\\
                        171 & Fe IX & quiet corona, upper transition region &$6.3\times10^5$\\
                        193 & Fe XII, XXIV &corona and hot flare plasma &$1.6\times10^6$, $2\times 10^7$\\
                        \hline
                \end{tabular}
        \end{center}
        \label{tables:aia}
\end{table*}

\section{Results}

\subsection{Quiet Sun region; single-dish ALMA}

Small-scale, bright ALMA features in the quiet region of the Sun observed by the ALMA, 
SDO-HMI and AIA, and GOES-SXI, instruments are shown in Fig.~\ref{Fig_1}. 
The quiet region of the Sun is denoted by the white rectangle in the full-disk ALMA image in Fig.~\ref{Fig_1}a. 

The four bright features marked by numbers in the lower left part of the selected quiet Sun region shown in Fig.~\ref{Fig_1}c are compared with a corresponding magnetogram, EUV, and soft X-ray images in panels d, e, and f, respectively.

It can be seen that ALMA bright features correspond very well with the most prominent magnetogram features. The features marked with 1, 2, 3, and 4 are all bipolar (or rather multipolar), and they also appear bright in coronal emissions. 
Feature 1, as the most probable case for CBP, is analyzed in detail in the bottom panels of Fig.~\ref{Fig_1} (g, h, i, and j). This feature lies over a large bipolar magnetic structure with an additional smaller patch of negative flux to the north-east. Emission in EUV shows several bright loops covering the region in the direction of the underlying magnetic flux. Similarly,  an enhanced emission is also visible in soft X-rays, but it is slightly  less compact, probably due to the lower image resolution. We can conclude that feature 1 is a CBP. Features 2 and 3 are less clear. Feature 2 is located above  a complex bipolar region, but it only appears as a bright diffuse region in EUV and soft X-rays, without any obvious structure. A small loop is visible to the north in 
193\,{\AA}, right above the magnetic bipole. 
 The X-ray emission from CBPs is emitted from the CBP loop tops, while the loop footpoints are seen in transition region temperatures. 
Feature 3 also lies over the complex bipolar region, but EUV emission only shows a small loop connecting a few magnetic flux concentrations of the opposite polarity. Again, in soft X-rays, two enhancements are barely visible, roughly corresponding to the footpoints of the loop. In conclusion, features 2 and 3 both also seem to be CBPs, but not as definitively as feature 1. Finally, feature 4 is not as bright as  the previous three features in 
 the ALMA image, but it shows significant emission in both EUV and soft X-rays. In  the AIA 193~\AA~ channel, a bright loop is visible arching over the locations of opposite polarity magnetic concentrations. It is most probably another case of CBP.
Other small-scale bright ALMA features
are mostly associated with unipolar magnetic field and are invisible in coronal emissions. 
They are most likely  related to solar plages or magnetic network elements.

\subsection{Active region; interferometric mosaic ALMA}

\subsubsection{A comparison of small-scale ALMA bright features with structures at other wavelengths}

\begin{figure*}
        \centering
        \includegraphics[width=19cm, trim={0 1.2cm 0 0},clip]{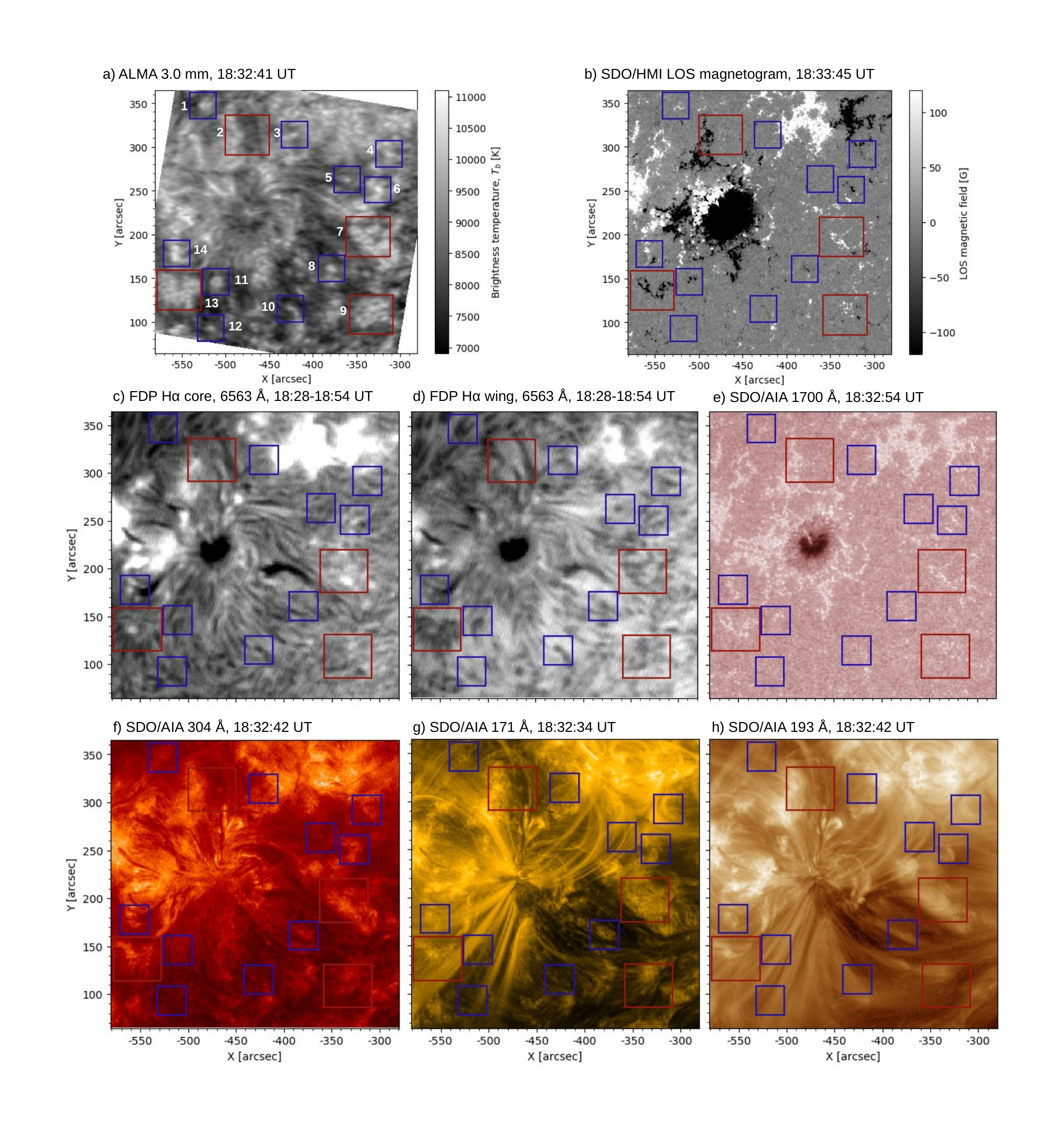}
        \caption{
        (a): Interferometric image of the solar active region 12470 obtained on 16 December 2015 at 100 GHz (3.0 mm) in ALMA Band 3. 
        The calibrated ALMA image has a field of view of 300" $\times$ 300". 
        The beam size of ALMA is 4.9" $\times$ 2.2". 
        (b) - (h): Images at other wavelengths taken close to the ALMA reference time (18:32 UT). 
        Squares lie at the same positions in all images, denoting selected  small-scale ALMA bright features.
        }
        \label{Fig_2}%
\end{figure*}

 We began by identifying the small-scale bright features in the ALMA image and checking their correspondence 
in images at other wavelengths.
A high-resolution interferometric ALMA image of the active region 12470 is presented in Fig.~\ref{Fig_2} 
together with the H$\alpha$ core and red and blue wing sum images from NSO-SOLIS-FDP and 
four SDO images in 171\,{\AA}, 193\,{\AA}, 304\,{\AA}, and 1700\,{\AA}, as indicated in the figure. 
We note that the beam size is almost an order of magnitude (7.3 times) smaller for the interferometric measurements than 
for the single-dish measurements. 
 In Fig.~\ref{Fig_2}a, several small-scale ALMA bright features were identified and denoted with squares. 
Smaller features are framed with smaller, blue squares, 
while larger features, showing different forms, lie within larger, red squares. All squares are enumerated
        and are located at the same positions in all other panels of Fig.~\ref{Fig_2}. 

 We compared visibility and shape of small-scale ALMA bright features 
with counterparts in other wavelength ranges. 
        The results  of this visual inspection are listed in Table~\ref{table:vis}. 
        In the second column, titled ALMA, we describe the shape of the ALMA bright
        feature for 14 selected areas. In the third column (HMI), we describe the characteristics of the line-of-sight (LOS) magnetic field,
        where {\em \emph{"complex"}} refers to shape and {\em \emph{"neutral"}}     denotes a very weak LOS magnetic field. 
        While the magnetic polarity is taken from the HMI single frame, 
        the dynamical evolution of magnetic structures is analyzed using multiple HMI frames in JHelioviewer, and 
        the information obtained this way complements the characterization in the third column.
        In the fourth and fifth columns,
        we analyze if the shape of the feature in the H$\alpha$ line matches that seen in ALMA and if the feature is dark or bright.
        In the sixth column, we look for matching shapes in SDO 1700\,{\AA} and ALMA, while in the next three columns, we list
        the visibility of the ALMA feature in SDO 304\,{\AA}, 171\,{\AA}, and 193\,{\AA} channels. The three SDO columns 
        are sorted in ascending height order according to Table~\ref{tables:aia}. 
        In the last column, we provide the  most plausible type of the  solar
        feature based on the observations  at other wavelengths.

        The LOS magnetic field varies in shape and polarity in the selected areas, indicating that the observed
        features are not of the same type. 
        A correlation between ALMA and the H$\alpha$ core exists, but it is far from unambiguous. Most bright ALMA features correspond to dark and 
        sometimes bright H$\alpha$ areas with a weak or good shape match. In the H$\alpha$ wing 
         (the sum of the blue and red wings of H$\alpha$), 
        the correlation is better, and the majority of ALMA 
        bright features         correspond to dark      areas. 
        A comparison of SDO 1700\,{\AA} and ALMA images in general shows weak correlation. 
         Shape matching  is very good in some cases,
         but for others it is nonexistent. In  the SDO 304\,{\AA}
        channel, we were able to identify all but one ALMA feature (ID = 12), while in 171\,{\AA} images, identification proved more difficult.
        This is the case for the 193\,{\AA} image, and about 50\% of features are barely visible, if at all.
        The analysis of 304\,{\AA}, 171\,{\AA}, and 193\,{\AA} SDO channels shows that the visibility systematically drops with increasing height
        and characteristic temperature (Table~\ref{tables:aia}).

\begin{figure*}[ht]
        \centering
        \begin{adjustbox}{addcode={\begin{minipage}{\width}}{\caption{Several selected ALMA regions, indicated by white rectangles in the first column images, shown enlarged in subsequent images at various wavelenghts: ALMA 3.0 mm, SDO-HMI magnetogram, SOLIS-FDP H$\alpha$ core 6563~\AA, SDO-AIA 304 and 193~\AA, respectively. ALMA contours are overlaid in each image. Contour levels are at 9000, 9500, 9750, 10000, 10250 and 10500 K.  IDs of selected regions are given in the first column.}
        \label{Fig_3}%
        \end{minipage}},rotate=90}
        \includegraphics[scale=1.02]{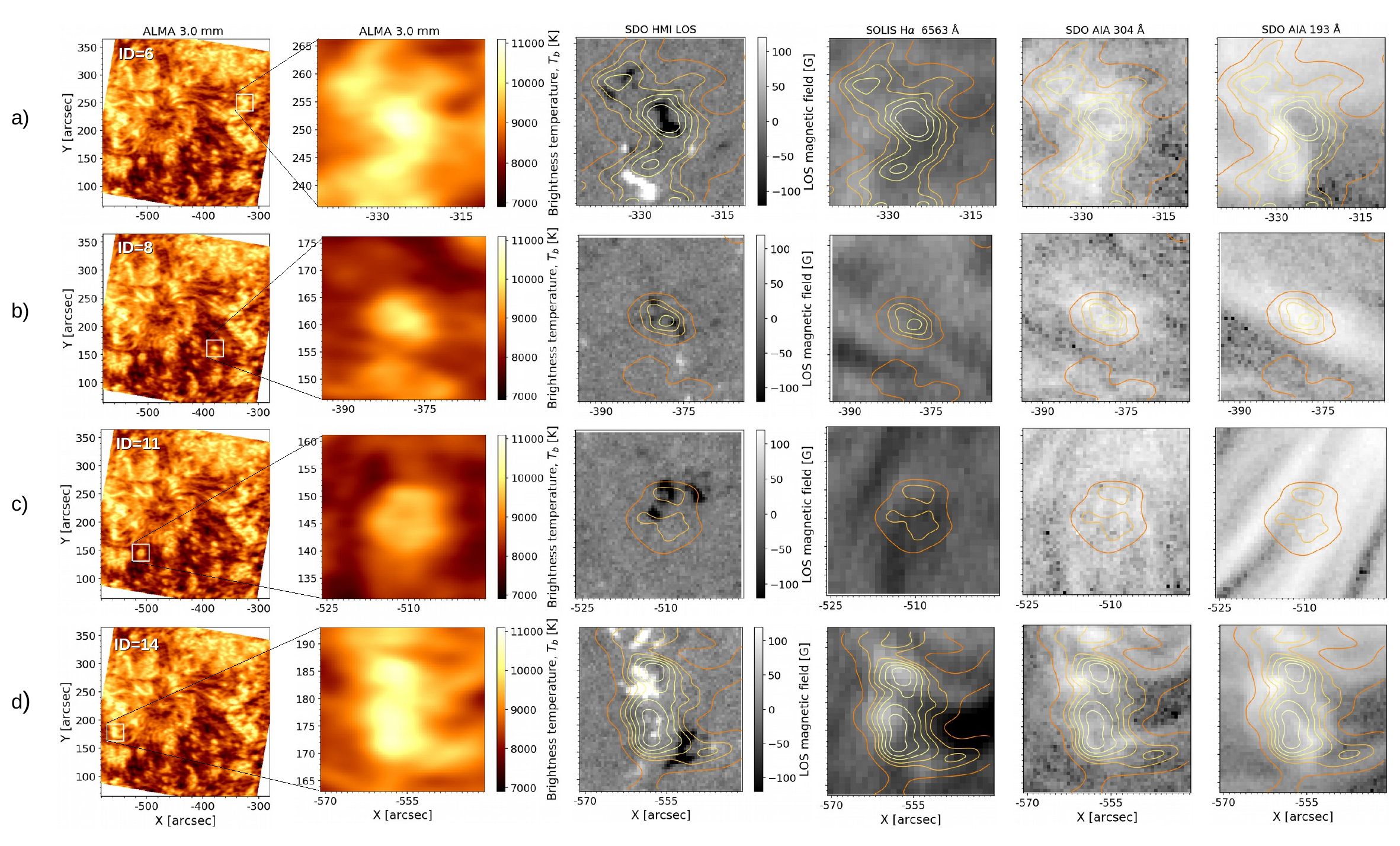}%
        \end{adjustbox}
                
        \end{figure*}

In Figs.~\ref{Fig_3}a-d, we show four regions of interest from Fig.~\ref{Fig_2} in more detail and compare them with  SDO-HMI, H$\alpha$ core, and SDO-AIA 304\,{\AA} and 193\,{\AA\ zoomed-in
        images}.  The ALMA contours are overlaid over all other images.
        In general, ALMA bright features are concentrated around
        strong LOS magnetic fields. 
        Correlation
        with the SDO 304\,{\AA} channel is very good, while it is worse in 193\,{\AA}, but still not bad. Again, one notable exception is 
         Fig.~\ref{Fig_3}c (ID = 11), 
        where ALMA is also bright in areas  in which the corresponding SDO-AIA images are not.
         The detailed analysis of these four small-scale bright ALMA features is also taken into account to resolve some ambiguous cases in 
        Table~\ref{table:vis}.  

\subsubsection{A comparison of identified H$\alpha$ structures with ALMA counterparts}

To further investigate a complex relationship between the ALMA and H$\alpha$ images of an active region, 
we searched for the ALMA counterparts of the known and identified H$\alpha$ structures. 
So, Fig.~\ref{Fig_4} shows a comparison between the H$\alpha$ core image and the ALMA interferometric 
image of the active region with H$\alpha$ core contours overlaid (upper panels). We then repeated the procedure using the H$\alpha$ wing sum 
image (lower panels). 
In the H$\alpha$ image, we analyze the most prominent dark areas (a sunspot and two small elongated filaments) 
and bright areas (four plages) in detail. These include the following: (i) the sunspot S located at $x = [-490", -460"]$, $y = [200", 240"], $  which appears dark in H$\alpha$ as well as at 3 mm 
(within the umbra, however, there is a local enhancement of the 3 mm radiation clearly visible in the ALMA image, without a notable 
H$\alpha$ counterpart); (ii) two small, dark H$\alpha$ filaments are visible to the left (FIL1) and to the right (FIL2) of the sunspot S at
$x_1 = [-560", -520"]$, $y_1 = [160", 190"]$ and 
$x_2 = [-420", -370"]$, $y_2 = [180", 200"]$. 
They are also clearly visible as dark structures in the 3 mm ALMA image; (iii) four bright plage areas, P1, P2, P3, and P4, are located at 
$x_1 = [-450", -270"]$, $y_1 = [290", 360"]$; 
$x_2 = [-580", -500"]$, $y_2 = [200", 250"]$;  
$x_3 = [-350", -310"]$, $y_3 = [170", 230"]$; 
$x_4 = [-330", -270"]$, and $y_4 = [80", 130"]$. 
All of them also appear bright in the ALMA image at 3 mm, as expected.

The H$\alpha$ wing sum image is more fragmented in comparison with the H$\alpha$ core image. 
The general appearance in ALMA is partly similar to the H$\alpha$ core, and the qualitative correspondence 
(bright-bright, dark-dark) can be seen in some cases. However, there are also some opposite examples, where 
dark areas in the H$\alpha$ wing sum correspond to bright areas in ALMA.

\section{Discussion}

\subsection{Quiet Sun region and single-dish ALMA}

Enhanced quiet Sun emission seen in ALMA is almost always associated with a strong LOS magnetic field. Moreover, enhanced ALMA emission in the single-dish ALMA image (Fig.~\ref{Fig_1})
follows the magnetic network very well, which can be seen from the good match between ALMA images, HMI magnetograms, 
SDO-AIA 193\,{\AA} channel  images, and soft X-ray GOES images. 
We selected a region of interest in the full-disk ALMA image, and within it we identified four possible cases of CBPs, 
while other small-scale ALMA bright features are most likely 
magnetic network elements and plages. 
One CBP from the quiet Sun visible in the ALMA image is analyzed in detail. 
Its elongated body clearly lies over a bipolar magnetic region in the photosphere (SDO-HMI magnetogram) 
and is strongly correlated with the CBP visible in emission  in EUV 193\,{\AA} and in soft X-rays.

\subsection{Active region and interferometric mosaic ALMA}

\subsubsection{A comparison of small-scale ALMA bright features with structures at other wavelengths}

The high-resolution ALMA image of an active region (Fig.~\ref{Fig_2}) reveals many small-scale bright features around the sunspot. We analyzed ten smaller,  mostly point-like structures and four larger, extended structures in more detail. 
 Generally speaking, small-scale bright  ALMA features are almost always associated with a strong
        LOS magnetic field in HMI images, frequently preserving the shape visible in ALMA.

        Images taken in SDO-AIA 304\,{\AA}, 171\,{\AA}, and 193\,{\AA} show decreasingly lower correlation with ALMA
        bright features, as these channels are sensitive to increasing heights and temperatures (Table~\ref{tables:aia}).
        The cause is probably linked to large magnetic loops overlying the active region (Fig.~\ref{Fig_2}h).
        As the loops become more visible and brighter, the contrast against lower lying features is lost, and bright,
        lower lying features are blended with the radiation coming from the large loops in EUV.

Using data from Table~\ref{table:vis}, which were composed by carefully inspecting Figs.~\ref{Fig_2} and \ref{Fig_3} and by checking the 
magnetic evolution of the structures on many subsequent frames using JHelioviewer, we can designate the most 
probable type of all of the14 identified small-scale bright ALMA features. 

Four cases are obviously CBPs (ID = 6, ID = 8, ID = 10, ID = 12) and one additional a possible candidate for the CBP (ID = 14). In all of these 
cases, the magnetic polarity is bipolar, with either emerging or canceling flux. In chromospheric and coronal lines, these features can be 
identified, with 
the characterization of "match" or "weak match", except in one case (ID = 12), where bright, coronal, large-scale structures overlay the CBP. 

There are two good plage candidates (ID = 2, ID = 7), which are magnetically unipolar and stable, and clearly bright in the H$\alpha$ core. 
One further case is ambiguous: it is either a CBP or a plage (ID = 9). The magnetic structure is bipolar, but complex, the feature is 
clearly bright in the H$\alpha$ core, and the correspondence with the chromospheric and coronal lines is good. 

In the next subgroup, we have five possible fibril candidates (ID = 1, ID = 4, ID = 5, ID = 11, ID = 13). The magnetic structure is mostly 
unipolar and stable, except in one case (bipolar, stable, ID = 5). All of these structures appear dark in the H$\alpha$ core. 

Finally, we have one special case (ID = 3) in which the ambiguity is between a fibril and a jet. This structure is magnetically neutral and dark in 
the H$\alpha$ core, exposing a weak match in SDO-AIA 304\,{\AA}, a streak in 171\,{\AA}, and no match in 193\,{\AA}. 

The correlation between small-scale ALMA bright features and the H$\alpha$ wing sum dark features is
        good (all cases except ID = 6, where the feature appears bright), although the shape is not completely preserved, which could be
        a consequence of lower resolution. 
        Dark fibrils in the H$\alpha$ wing might mark horizontal field lines connecting opposite magnetic polarities
        \citep{Zirin1988, Madjarska2020}. 
        This is in agreement with a high correlation of ALMA bright points and magnetic structures, although in some cases only one polarity could be related to the mm and 
        H$\alpha$ structure. We also note that fibrils might show significant magnetic evolution, as pointed out by \citet{Tandberg1995} and 
        described in more detail in the Introduction.

        The distance between opposite magnetic polarities for feature ID = 6 is $\approx$8500 km. 
        Assuming a semicircular shape of the loop, we can estimate the height to be $\approx$4000 km. This places the feature in the lower corona, supporting the idea that the object is a CBP, even though this height is lower than that typically found in other CBP studies \citep{Brajsa2004, Sudar2016}. \citet{Shimojo_Hudson2017} also detected a CBP in close proximity to a sunspot.
        We also note that this feature appears bright in the H$\alpha$ wing sum, although in all other cases CBPs appear dark in the 
        H$\alpha$ wing sum.
 Using high-resolution EUV and H$\alpha$ images, \citet{Madjarska2020} recently studied the chromospheric counterpart of the CBP  and found that it is composed of a bundle of dark elongated features 
called H$\alpha$ loops. The footpoints of these H$\alpha$ loops are rooted in the magnetic flux concentrations  of opposite polarity and appear 
as bright features in H$\alpha$. 
The brightenings have sharp edges in the line wings, while they appear larger and more diffuse at the edges in the line center.

Based on the characteristics of small-scale bright ALMA features summarized in Table~\ref{table:vis}, we now present joint statistics of the 
distribution of various properties found, with the number of cases in parenthesis. We note that there are always 14 structures in total. 
According to the shape in ALMA we found: multiple points (3), complex (7), and point-like structures (4).
Regarding the magnetic polarity according to HMI we found: unipolar (6), bipolar (7), and neutral (1).
Regarding the magnetic structure according to JHelioviewer we found: stable (6), canceling flux (5), emerging flux (2), 
and N/A (1)\footnote{This one case 
refers to the neutral magnetic polarity feature, in which no structure and dynamics could be determined.}. 
According to the visibility in the H$\alpha$ core we identified: 
dark (9): match (5) and weak match (4); bright (4): match (1) and weak match (3); no match (1). 
According to the visibility in the H$\alpha$ wing sum we identified:
dark (13): match (8) and weak match (5); bright (1): weak match (1).
Furthermore, regarding the visibility in SDO-AIA 1700~\AA\, we found: match (7), weak match (2), and no match (5).
Regarding the visibility in SDO-AIA 304~\AA\, we found: match (8), weak match (5), and no match (1).
According to the visibility in SDO-AIA 171~\AA\, we found: match (5), weak match (5), streak (1), and no match (3).
According to the visibility in SDO-AIA 193~\AA\, we found: match (1), weak match (7), and no match (6). 
Finally, regarding the type of structures we concluded: CBP (5), plage (2), fibril (5), jet? or fibril? (1), and CBP? or plage? (1).

\subsubsection{A comparison of identified H$\alpha$ structures with ALMA counterparts}

\begin{figure*}
                \centering
                \includegraphics[width=19cm, trim={0 0.5cm 0 0},clip]{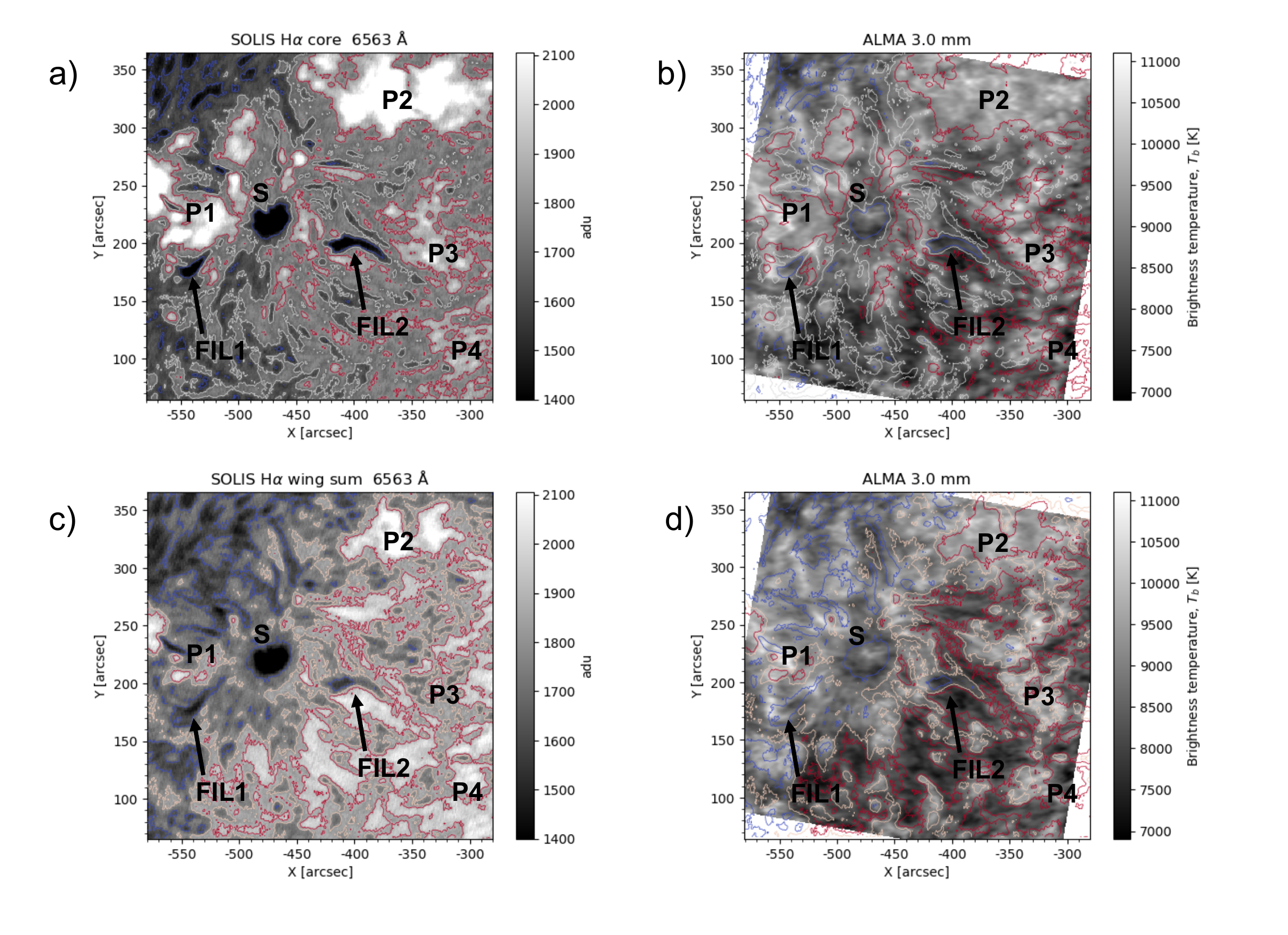}
                \caption{ Comparison between (a) the H$\alpha$ core image  and (b): the ALMA image  with H$\alpha$ core contours overlaid, (c) the H$\alpha$ wing sum image,  and (d) the ALMA image  with H$\alpha$ wing sum contours overlaid. Blue, white, and red contours are in increasing intensity order.
                In panel (a) "P" denotes the four plages, "S" the sunspot, and "FIL" the two small filaments. 
                }
                \label{Fig_4}%
        \end{figure*}

A comparison between the H$\alpha$ core image and the interferometric 3 mm ALMA image of the analyzed active region with contours of 
H$\alpha$ overlaid over the ALMA image (Fig.~\ref{Fig_4}) reveals a good correspondence. The sunspot S appears dark in both images 
(with a local ALMA radiation enhancement in the sunspot umbra), the four 
plage areas, P1, P2, P3, and P4, appear bright in both images, and the two dark H$\alpha$ filaments (FIL1 and FIL2) are also clearly recognized as dark 
structures of the same shape in ALMA. 
It is interesting that while FIL1 lies along the magnetic inversion line, as can be seen in the LOS magnetogram (Fig.~\ref{Fig_2}), FIL2 is not associated with any discernible LOS magnetic structure. 
We also checked the evolution of the FIL2 in a series of H$\alpha$ images from the Solar Monitor\footnote{http://solarmonitor.org}, 
eight hours earlier and 15 hours 
later than the time of the ALMA image. The small filament FIL2 shows a clear development growing in size during the 24-hour period 
under consideration.

A comparison between the H$\alpha$ wing sum image and the interferometric 3 mm ALMA image of the analyzed active region with contours of 
H$\alpha$ overlaid over the ALMA image (Fig.~\ref{Fig_4}d) reveals a more complicated relationship than for the H$\alpha$ core case. 
The sunspot is also dark in the H$\alpha$ wing sum image, as expected. 
Both filaments are dark in all three images: H$\alpha$ core, H$\alpha$ wing sum, and ALMA. 
The plage areas P1 and P2 are bright in all three images: H$\alpha$ core, H$\alpha$ wing sum, and ALMA.
However, the plage area P3 is bright in the H$\alpha$ core, dark in the H$\alpha$ wing sum, and bright in ALMA. 
Furthermore, the
plage area P4 has a less defined bright shape in the H$\alpha$ core, corresponding to a less defined shape (partly bright and partly
dark) in the H$\alpha$ wing sum, while the ALMA shape corresponds to the
brightness-inverted wing sum shape. Finally, many
small-scale structures that are bright in the H$\alpha$ wing sum are
dark in ALMA and vice-versa.
At present, we do 
not have a satisfactory explanation for this curious relationship between the H$\alpha$ wing sum and ALMA images, but we note that the most 
significant difference in comparison to the H$\alpha$ core image appears in the case of plages P3 and P4.

\section{Summary}

In this work, we analyzed small-scale ALMA bright features in the quiet Sun region using single-dish ALMA observations (1.21 mm, 248 GHz) and 
in an active region using interferometric ALMA measurements (3 mm, 100 GHz). 
Using  single-dish observations, a full-disk solar image was produced, while interferometric measurements enable the high-resolution
reconstruction of a  part of the solar disk, including an active region. 
The selected quiet Sun region was compared with the SDO-HMI magnetogram, with the SDO-AIA 193~\AA\, image, and with the soft X-ray 
GOES image. 
The ALMA image of the selected active region was compared with a series of other images: 
SDO-HMI magnetogram,  H$\alpha$ core and wing sum, SDO-AIA 1700~\AA, 304~\AA, 171~\AA, and  
193~\AA.  The dynamics of the magnetic structure was inspected by SDO-HMI videos from JHelioviewer, and the quality and 
resolution of the H$\alpha$ images were improved using the stacking procedure. The results are summarized in Table \ref{tables:vis}, where the 
visibility of the selected small-scale ALMA bright features is checked in other images. 
Finally, we also compared the ALMA image of the active region with the corresponding H$\alpha$ core and wing sum images, using   
contour overlapping, to check the ALMA correspondence of various known chromospheric structures visible in the 
H$\alpha$ images of the Sun.


Based on the analysis and results presented in this work, we can conclude that in the quiet Sun region,
enhanced emission seen in ALMA is almost always associated with a strong LOS magnetic field. 
In the selected quiet Sun region, four good chromospheric correspondences for CBPs were identified, while other small-scale ALMA bright features are most likely associated with magnetic network elements and plages.

In the active region 14 small-scale ALMA bright features were randomly selected, compared with other images, and analyzed in detail.
For CBPs, we found five good candidates. In all of these cases, the magnetic polarity is bipolar, with either emerging or canceling flux.
Except in one case, there is a good correspondence with structures in chromospheric and coronal lines. 
In four cases, we found a good correspondence with H$\alpha$ wing sum dark features, and in one case with a bright feature. 
It is interesting to note that \citet{Madjarska2020} identified the chromospheric counterpart of the CBPs  as dark elongated features 
called H$\alpha$ loops, which correspond to dark H$\alpha$ fibrils, but the footpoints of these H$\alpha$ loops are rooted in the magnetic flux concentrations  of opposite polarity and appear as bright features in H$\alpha$. 
There are two good candidates for plages, which are magnetically unipolar and stable and clearly bright in the H$\alpha$ core. 
One further case is ambiguous: it is either a CBP or a plage. 
We found five possible fibril candidates. Their  magnetic structure is mostly unipolar and stable, except in one
case where it is bipolar. All of these structures appear dark in the H$\alpha$ core.
We note that the identification of fibrils as small-scale bright ALMA features is less certain than in case of CBPs. 
Finally, there are two unclear cases. In the first, it is unclear if the ALMA feature is a fibril or a jet. The structure is magnetically neutral, dark in H$\alpha$ core, 
and hidden by loop-like streaks in upper transition region and coronal lines.
In the second ambiguous case, it remains unclear if the ALMA feature is a CBP or a plage. The magnetic structure is bipolar, but 
complex, the feature is clearly bright in the H$\alpha$ core, and the correspondence with the chromospheric and coronal lines is good. 
The correlation between small-scale ALMA bright features and H$\alpha$ wing sum dark features is very high, as it  was 
confirmed in all but one case. Dark fibrils in the H$\alpha$ wing might mark horizontal field lines connecting opposite magnetic polarities 
 \citep{Zirin1988, Madjarska2020}, but the fibrils might also show significant dynamics and magnetic evolution, as pointed out by 
\citet{Tandberg1995}. 

A comparison of the H$\alpha$ core image and the ALMA image of the analyzed active region with contours of 
H$\alpha$ overlaid over the ALMA image showed that the sunspot appears dark in both images 
(with a local ALMA radiation enhancement in sunspot umbra), the four 
plage areas bright in both images, and small dark H$\alpha$ filaments, are also clearly recognized as dark structures of the same shape in ALMA. 
Finally, a comparison of the H$\alpha$ wing sum image and the ALMA image of the analyzed active region shows a similar behavior to that of 
the H$\alpha$ core case, with the exception that two plage areas are completely or partly dark in the H$\alpha$ wing sum image. 
Moreover, many small-scale structures, which are bright in the H$\alpha$ wing sum, are dark in ALMA and vice-versa, implying a bright-dark 
and dark-bright correlation.

\section{Conclusions}

For the first time, a correspondence between small-scale ALMA bright features and  various chromospheric and coronal structures, 
CBPs, plages, fibrils, and jets, was found. 
Most of the ALMA structures originate in the chromosphere coupled with regions of strong photospheric magnetic field. 
Some smaller structures with higher temperatures 
might be at higher altitudes, that is in the transition region.

 We note that a  generally good one-to-one correspondence of the structure's positions and shapes in the ALMA image with other images 
was found. This confirms that the reconstruction of the ALMA interferometric image and the overlay of various images (orientation, co-alignment) has been done accurately and reliably  and also demonstrates that ALMA features are real and not image synthesis artifacts.
We conclude that small-scale bright ALMA features with soft X-ray enhancements 
in the quiet Sun region trace the locations of CBPs and the underlying magnetic network.

A general conclusion for the active region is that ALMA small-scale bright structures follow magnetic structures from SDO-HMI and 
SDO-AIA 1700~\AA\, and 304~\AA\, bright features rather well. 
This leads to the conclusion that bulk of the millimeter radiation originates from the chromosphere and transition region (Table \ref{tables:aia}) 
and from the structures above areas of the increased photospheric magnetic field, as can be seen in the SDO-HMI 
images. This is in agreement with the fact that the ALMA Band 3 originates from higher chromospheric levels, 
and the ALMA Band 6 originates from lower ones. 
Some small-scale ALMA structures are also discernible in the SDO-AIA 171~\AA\, and 193~\AA\, channels. 
Lower chromospheric structures can be partly seen with ALMA through the optically thin corona, whereas SDO-AIA 193~\AA\, 
loops are not visible.

The identification of small-scale ALMA bright features in active regions as EUV CBPs might be complicated in some cases. 
The cause is probably the appearance of  large bright magnetic loops overlying the active region and in this way 
        the contrast against lower lying features is lost, and bright,
        lower lying features are blended with the radiation coming from the large loops in EUV.

\citet{Rutten2017III} predicted that the general appearance of the solar chromosphere at the 
ALMA wavelengths will be similar to the H$\alpha$ core images with a good dark-dark correspondence. 
A good dark-dark correspondence (sunspot and small filaments) and bright-bright correspondence (plages)  
between the two images was found by the present analysis. However,  
the relationship between H$\alpha$ core and high-resolution ALMA images in the millimeter wavelength range 
is more complicated, since we find examples of both types of behavior: some small-scale, bright ALMA features have bright counterparts, and some have 
dark H$\alpha$ counterparts. 
Moreover, we also found examples of a bright-dark and a dark-bright correlation between the H$\alpha$ wing sum and ALMA images. 

Finally, a local enhancement of 3 mm ALMA radiation is visible within the sunspot umbra, confirming an 
earlier analysis of the same active region by \citet{Iwai2017}. 
This umbral brightening observed with ALMA can also be recognized in the SDO-AIA 304~\AA\, and 171~\AA\, channels, indicating 
a transition region height.

Increased ALMA brightness suggests a higher temperature in the emission region where the magnetic field is strong. Because of the high correspondence of ALMA bright features with magnetic flux elements, future theoretical modeling of small-scale ALMA bright features should also include detailed magnetic structure.
We note that a detailed magnetic topological analysis of CBPs was performed by 
\citet{Galsgaard2017}. In the present analysis, we tried the usual potential field source surface (PFSS) model calculation for both the 
quiet Sun and active regions, but this did not render sufficient new information, because the spatial resolution of  the procedure we used was too low. 
So, we leave this type of analysis to be done in more detail in a later work. 
A better constraint on the height range of the observed structures will be possible using a tomographic method when additional observing 
frequencies (ALMA observing bands) become available. This is also left for future research.

\begin{acknowledgements}

This work has been supported by the Croatian Science
Foundation under the project 7549 "Millimeter and submillimeter
observations of the solar chromosphere with ALMA". 

It has also received
funding from the Horizon 2020 project SOLARNET (824135, 2019--2022). 

In this paper ALMA data ADS/JAO.ALMA\#2011.0.00020.SV were used. 
ALMA is a partnership of ESO (representing its member
states), NSF (USA) and NINS (Japan), together with NRC (Canada), MOST and ASIAA (Taiwan), 
and KASI (Republic of Korea), in cooperation
with the Republic of Chile. The Joint ALMA Observatory is operated by ESO, AUI/NRAO and NAOJ.
We are grateful to the ALMA project for making solar observing with ALMA possible.

SDO is the first mission launched for NASA\'\,s Living With a Star (LWS) Program.

This work utilizes data obtained by the Global Oscillation Network Group (GONG) Program, managed by the National Solar Observatory (NSO),
which is operated by AURA, Inc. under a cooperative agreement with the National Science Foundation (NSF). The data were acquired by instruments operated by the Big Bear Solar Observatory, High Altitude Observatory, Learmonth Solar Observatory, Udaipur Solar Observatory, Instituto de Astrofisica de Canarias, and Cerro Tololo Interamerican Observatory.
This work utilizes SOLIS data obtained by the NSO Integrated Synoptic Program (NISP), managed by the National Solar Observatory (NSO), 
which is operated by the Association of Universities for Research in Astronomy (AURA), Inc. under a cooperative agreement with the National Science Foundation (NSF).

RB acknowledges financial support from the Alexander von Humboldt Foundation. 
HGL acknowledges financial support by the Sonderforschungsbereich SFB\,881
"The Milky Way System" (subprojects A4) of the German Research
Foundation (DFG).
CLS acknowledges financial support from the S{\~a}o Paulo Research Foundation (FAPESP), grant number 2019/03301-8.

We thank Manuela Temmer and Bojan Vr\v snak for helpful remarks and discussion. 

Finally, we thank the anonymous referee for insightful comments and suggestions which led to the significantly improved paper.

\end{acknowledgements}

\begin{table*}
        \caption{ Visibility of the selected small-scale ALMA bright features from Fig.~\ref{Fig_2}a in other wavelengths.
        }
        \label{table:vis}      
        \centering                          
        \begin{tabular}{|m{1cm}|m{1.3cm}|m{1.5cm}|m{1.5cm}|m{1.5cm}|m{1.3cm}|m{1.3cm}|m{1.3cm}|m{1.3cm}|m{1.3cm}|}
                \hline\hline                 
Feature ID & Shape in ALMA & HMI, JHelioviewer & H$\alpha$ core & H$\alpha$ wing sum & SDO-AIA 1700~\AA & SDO-AIA 304~\AA & SDO-AIA 171~\AA & SDO-AIA 193~\AA & Type \\    
                \hline                        
1 & multiple points & unipolar, stable & dark, weak match & dark, weak match & weak match & weak match & weak match & no & fibril  \\ \hline

2 & complex & unipolar,  stable & bright, weak match & dark, weak match & match & match & match & weak match & plage \\ \hline

3 & point & neutral & dark, match & dark, match & no & weak match & streak & no &  jet?, fibril? \\ \hline

4 & multiple points & mostly unipolar, cancelling flux at the edges & dark, match & dark, match & match & match & weak match & no & fibril  \\ \hline

5 & point & small, bipolar, stable & dark, match & dark, match & no & weak match & no & no & fibril \\ \hline

6 & complex & bipolar, emerging flux & no match & bright, weak match & match & match & weak match & weak match & CBP \\ \hline

7 & complex & unipolar,  stable & bright, match & dark, match & match & match & match & weak match & plage  \\ \hline

8 & point & bipolar, cancelling flux & dark, match & dark, match & no & match & match & weak match & CBP \\ \hline

9 & complex & bipolar, complex, cancelling flux at the edges, dynamic & bright, weak match & dark, weak match & weak match & match & match & weak match & CBP?, plage? \\ \hline

10 & multiple points & bipolar, cancelling flux & dark, match & dark, match & no & weak match & weak match & weak match & CBP \\ \hline

11 & complex & unipolar, stable & dark, weak match & dark, weak match & match & weak match & no & no & fibril  \\ \hline

12 & point & weak bipolar, cancelling flux & dark, weak match & dark, match & no & no & no & no & CBP \\ \hline

13 & complex & unipolar, stable & dark, weak match & dark, match & match & match & match & weak match & fibril \\ \hline

14 & complex & bipolar, emerging flux & bright, weak match & dark, weak match & match & match & weak match & match &  CBP \\          
                \hline                                   
        \end{tabular}
        \label{tables:vis}      
\end{table*}

\bibliographystyle{aa}
\bibliography{36231corr} 


\end{document}